# Anatomy of the Sagittarius A complex:

## IV. Sgr A* and the Central Cavity revisited

R. Zylka[1], P.G. Mezger[1], D. Ward-Thompson[2],
W.J. Duschl[1,3], and H. Lesch[1]

1: Max-Planck-Institut für Radioastronomie, Auf dem Hügel 69, D-53121 Bonn, Germany
2: Royal Observatory, Blackford Hill, Edinburgh, EH9 3HJ, Scotland, UK
3: Institut für Theoretische Astrophysik, Ruprecht-Karls-Universität,
   Im Neuenheimer Feld 561, D-69120 Heidelberg, Germany

# Abstract

We present submm images of Sgr A* and its surroundings obtained at $\lambda 800$, 600 and $450\mu$m with the James Clerk Maxwell Telescope, JCMT[1], and derive flux densities of Sgr A* at all three wavelengths. Combined with upper limits by Gezari and associates at MIR wavelengths a time averaged radio spectrum is obtained which increases $\propto \nu^{1/3}$, attains a maximum at $\nu_{max} \gtrsim 600\,$GHz and must decrease rapidly at frequencies $\gtrsim 10^4$GHz. This spectrum allows for about $3\,M_\odot$ of 50K dust and associated hydrogen in the telescope beam. While variations on time-scales of a few months are now well established for the frequency range $\nu \lesssim 100\,$GHz, our investigation of variability at higher frequencies still yields only marginal results.

The Circum-Nuclear Disk (CND) extends over the central 12 pc. At the galactocentric radius $R \sim 1$ pc dust and hydrogen column densities drop to low values and form the Central Cavity. Our submm images show that the bottom of this cavity is rather flat. Variations in the dust emission are just consistent with the detection of a 'Tongue' of $\sim 200\,M_\odot$ of atomic hydrogen reported by Jackson et al. (1993) to be located between the Northern and the Eastern Arm of the Minispiral. We present a revised submm/IR spectrum of the central 30″ ($R \lesssim 0.6$ pc) with flux densities corrected for an interstellar extinction of $A_v \sim 31$ mag. This spectrum attains its maximum at $\lambda \sim 20\,\mu$m and comes from dust with temperatures $\sim 170-400$ K which is associated with the Eastern Arm and the East-West Bar. The integrated luminosity is $\sim 5\,10^6\,L_\odot$ to which emission at $\lambda \lesssim 30\,\mu$m contributes $\sim 80\%$.

Heating of this dust is not provided by a central source but rather by a cluster of hot ($T_{eff} \sim 3-3.5\,10^4$ K) and luminous stars which could include the HeI/HI-stars detected by Krabbe et al. (1991). Their total luminosity within $R \lesssim 0.6$ pc must be $\sim 2\,10^7\,L_\odot$ and their Lyman continuum photon production rate $N_{Lyc} \sim 1.3\,10^{51}\,s^{-1}$. Outside the Central Cavity the principal sources of excitation appear to be medium mass stars and late-type O-stars. IR- and Lyc-photon luminosities increase $\propto R^{1.2}$ while the dust temperature decreases $\propto R^{-0.15}$.

Within the Nuclear Bulge ($R \lesssim 250$ pc) the IR-luminosity ($\sim 10^9\,L_\odot$), the production rate of Lyman continuum photons ($\sim 2\,10^{52}\,s^{-1}$) and the mass of molecular hydrogen ($\sim 10^8\,M_\odot$) amount to $\sim 10\%$ of the corresponding characteristics in the Galactic Disk. Hence the formation rate of massive stars per unit mass of molecular hydrogen and the stellar population in Bulge and Disk must be comparable.

Comparison with model computations show that the formation of the Central Cavity is a consequence of the structure of the gravitational potential, which is dominated for $R \gtrsim 1$ pc by stars of the Nuclear Bulge and for $R \lesssim 1$ pc by a compact object of a few $10^6\,M_\odot$. Model computations for a black hole/accretion disk configuration with an accretion rate of $10^{-6...-7}\,M_\odot\,yr^{-1}$ indicate that the accretion disk becomes gravitationally unstable for $\gtrsim 1.5-5\,10^4$ Schwarzschild radii. A more massive disk with a correspondingly higher accretion rate thus could account for the formation of the central cluster of HeI/HI and – possibly – O-stars which appear to be responsible for most of the excitation of the central pc.

---

# 1 Introduction

Sgr A* is a compact synchrotron radio source located at or very close to the dynamical center of the Galaxy. It is believed to be a starved black hole of $\sim 2\,10^6\,M_\odot$ (see, e.g., Falcke et al. 1993). Its radio spectrum $\nu \lesssim 100\,\text{GHz}$ - referred to as $S_\nu(\text{A}^*,1)$ - is slightly inverted and varies on time scales of a few months (Zhao et al. 1991). In the short mm- and submm range the spectrum - referred to as $S_\nu(\text{A}^*,2)$ – rises more steeply (Zylka and Mezger 1988; Zylka, Mezger and Lesch 1992, hereafter referred to as ZML 92). No positive detections of Sgr A* at FIR and MIR wavelengths have been reported, but a series of significant upper limits in the wavelength range $18 \geq \lambda/\mu\text{m} \geq 8$ were obtained by D. Gezari (as quoted in ZML 92) which indicate that this part of the $S_\nu(\text{A}^*,2)$ spectrum must drop by more than an order of magnitude between the submm and the MIR regime.

ZML 92 offered two explanations for this spectral behavior. In both cases optically thin synchrotron emission from a compact source explains the slightly inverted radio spectrum $S_\nu(\text{A}^*,1)$. The steep rise of the spectrum $S_\nu(\text{A}^*,2)$ in the mm range was alternatively explained by i) a second even more compact component with self-absorbed synchrotron emission; or ii) thermal $30-60\,\text{K}$ dust emission from a compact ($\lambda_{\tau=1} = 1\,\text{mm}$) cloud containing a few hundred solar masses of gas and dust.

A specific subgroup of the first alternative are models based on the assumption of a very narrow energy distribution of the relativistic electrons, also referred to as 'Monoenergetic Electrons'. If their synchrotron radiation is optically thin, the spectrum increases with $S_\nu \propto \nu^{1/3}$, attains a maximum at $\nu_{max}$ and decreases for $\nu > \nu_{max}$ with $S_\nu \propto \exp(-\nu/\nu_c)$, with $\nu_c = 3\nu_{max}$ the cut-off frequency. Duschl and Lesch (1994a; hereafter: DL94) found that the time-averaged spectrum of Sgr A* can be approximated by such a monoenergetic spectrum. This model will also be discussed here in the context of an interpretation of the revised mm/IR spectrum of Sgr A*.

In a review of the characeristics of Sgr A* completed in summer 1993 Mezger (1994; hereafter referred to as M94) assembled all relevant observations available at that time. A set of measurements with the JCMT 15 m telescope by Dent et al. (1993) at four wavelengths ($\lambda 2200$, 2100, 800 and $450\,\mu\text{m}$) and by Serabyn and Lis (1994) at $\lambda 350\,\mu\text{m}$ constrained the ZML 92 model of thermal dust emission, especially since at the two shortest wavelengths only upper limits of $\leq 1.5$ and $\leq 10\,\text{Jy}$ respectively were obtained. However, all these observations have rather high uncertainties which arise mainly from two sources: Anomalous refraction (Altenhoff et al. 1987; Coulman 1991) and the fact that Sgr A* is observed against a background of free-free and dust emission. Since at submm wavelengths this background emission increases as $S_\nu \propto \nu^4$ while $S_\nu(\text{A}^*,2)$ remains rather flat the uncertainties in the Sgr A* flux densities increase rapidly with frequency and attain values at $\lambda \lesssim 450\mu\text{m}$ which are comparable with or even larger than the expected flux densities of Sgr A*. This explains the quite different upper limits quoted above for the adjacent bands at 450 and $350\,\mu\text{m}$ (see also M94).

The above mentioned review concluded that 'although some discrepancies need to be cleared, the high-resolution submm observations make the conclusion nearly unavoidable that the spectrum $S_\nu(\text{A}^*,2)$ levels off (and must turn over) at $\nu \geq 350\,\text{GHz}$. Alternatively, a high variability of the submm/FIR emission could be suggested'. Here we present new results on both the spectral behavior and time variability of $S_\nu(\text{A}^*,2)$ which lead to a rediscussion of the nature of Sgr A*.

The stellar environment appears to play an important role in the physics of the Galactic Center. While the Galactic Bulge ($R \lesssim 2.5\,\text{kpc}$) probably is a continuation of the halo and thus is populated by old Pop II stars, there appear to coexist within the central pc two different star populations: i) Older ($\sim 0.1\,\text{Gyr}$) medium mass stars which belong to the Nuclear Bulge ($R < 250\,\text{pc}$) and where M- and K-giants contribute most of the luminosity. These giants constitute only a minor fraction of the total stellar mass whose volume density increases with galactocentric radius $\propto R^{-1.8}$. The core radius of this cluster could be as large as $\sim 0.8\,\text{pc}$. ii) A cluster of younger ($\sim 10\,\text{Myr}$) massive and luminous stars of core radius $\sim 0.15\,\text{pc}$, the most conspicuous of which belong to the class of HeI/HI stars. These stars represent a short transition period of O-stars on their evolutionary path to WR-stars. More than a dozen of these stars have been identified within the central pc. The present detection limits do not exclude the presence of MS-O-stars in the inner 30″. (For recent reviews see Eckart et al. 1993; Genzel et al. 1994c; Allen 1994; Rieke and Rieke 1994; and Genzel et al. 1994b). Allen goes as far as to suggest that as a result of dynamical interaction the newly formed massive stars have pushed out any Pop II stars that were within the central pc.

Table 1: JCMT mapping parameters of the images shown in Fig. 1.

| $\lambda$ | HPBW | No.of maps | Map ext. | Pixel size | Chop. throw | Int. time per pixel | Zenith opacity at 230 GHz |
|---|---|---|---|---|---|---|---|
| $\mu$m | arcsec | | arcsec | arcsec | arcsec | sec | 230 GHz |
| 800 | 13 | 3 | 156×56 | 4×4 | 40 | 1 | 0.04 |
| | | 2 | 184×64 | 4×4 | 44 | 1 | 0.04 |
| 600 | 10 | 3 | 156×56 | 4×4 | 30 | 2 | 0.03 |
| | | 4 | 180×48 | 3×3 | 33/40 | 2 | 0.03 |
| 450 | 8 | 2 | 240×40 | 3×3 | 30 | 1 | 0.04 |
| | | 7 | 180×48 | 3×3 | 30 | 2 | 0.026 − 0.03 |

The detection of intense massive star formation in the immediate vicinity of a black hole comes as a surprise especially since these stars can supply most of the luminosity in the central pc. To further investigate this configuration we present a revised dereddened spectrum of the central parsec from which we determine the total IR luminosity $L_{IR}(30'')$ and the dust temperature.

For radii $R > 1$ pc the composition of stars appears to change drastically. We use the excitation of the interstellar matter – i.e., the ionization of the gas and the dust temperature – as a means to investigate the stellar composition of the Nuclear Bulge close to the Galactic plane. Combining our results with observations published elsewhere we determine $L_{IR}$ and the Lyman continuum photon production rate as a function of galactocentric radius $R$ and discuss the origin of the excitation – i.e., ionization of the gas and heating of the dust – from the vicinity of Sgr A* out to a galactocentric radius of $R \sim 250$ pc.

The configuration of the central 2 pc – i.e., Central Cavity and star cluster – is found to be consistent with the presence of a black hole surrounded by an accretion disk.

## 2 Observations and Results

### 2.1 Observations with the JCMT: The revised spectrum $S_\nu(\mathbf{A^*}, 2)$

The observations with the JCMT were performed during the second shift, $01^h30^m - 09^h30^m$ HST ($12^h30^m - 20^h30^m$ UT), of the nights of February 27 to March 1, 1994. Sgr A* and its surroundings were mapped at 800, 600 and 450 $\mu$m using the common-user receiver UKT14, which contains a single-element, ${}^3$He-cooled Ge:In:Sb bolometer, and a series of filters matched to the atmospheric transmission windows (Duncan et al. 1990). Raster mapping in azimuth-elevation was carried out in 'on-the-fly' mode, and the mapping parameters are listed in Table 1. The observations were carried out while using the secondary mirror to chop in azimuth at around 7 Hz and synchronously to detect the signal, thus rejecting 'sky' emission. Focus checks were carried out every $\sim 1 - 2$ hours. Weather conditions during the observations varied from very good to excellent, with zenith opacities at 230 GHz decreasing from 0.04 on Feb 27 to 0.026 on March 1 (230 GHz opacities were obtained from the NRAO radiometer located at the CalTech Submm Observatory).

The raw dual-beam maps were created using the JCMT task MAKEMAP. The remaining data reduction was performed with the software package MAP written by R. Zylka. The images were corrected for constant baseline offsets, calibrated and restored to single-beam maps using the modified algorithm of Emerson, Klein and Haslam (1979).

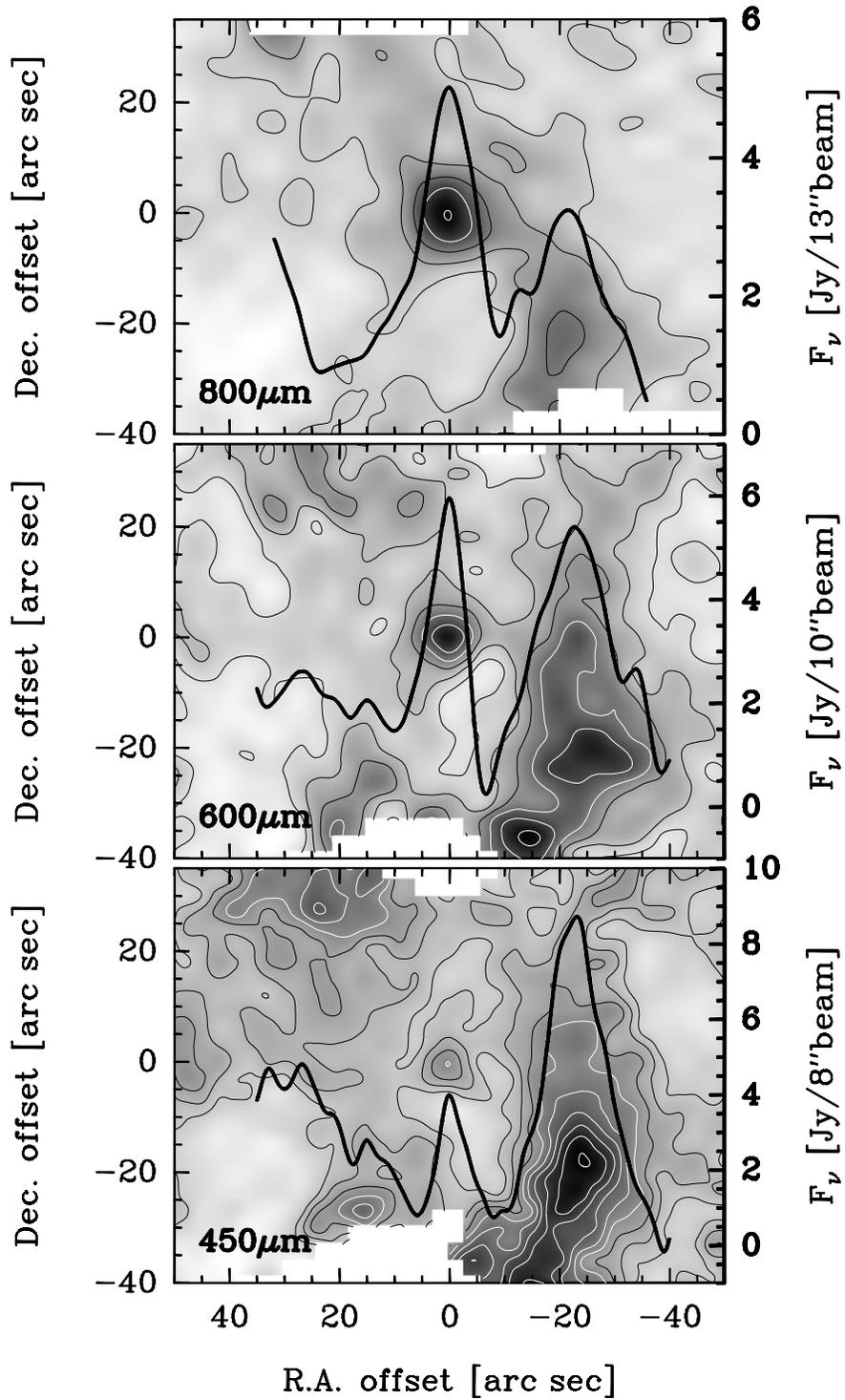

Figure 1: Averaged JCMT images of the central ∼ 2′ obtained with the UKT14 bolometer at λ800, 600 and 450 μm. Contours are drawn in equidistant steps of 1 Jy/beam beginning at 2 Jy/beam. Cuts through the images at a position angle of 45°, i.e. from SW to NE, are shown as black solid lines. The intensity scales for these cuts are indicated on the right image ordinate. $\Delta\alpha, \Delta\delta$ are coordinates offset relative to the position of Sgr A*, $\alpha = 17^{\mathrm{h}}42^{\mathrm{m}}29\overset{\mathrm{s}}{.}314$, $\delta = -28°59'18\overset{''}{.}3$ (epoch 1950, see, e.g., Rogers et al. 1994).

Anomalous refraction can be considered as the radio analogue of optical seeing but with beam displacements up to 40″. (For a discussion of this effect see Appendix A of ZML92 and Appendix B of this paper). To minimize the effects of anomalous refraction the highest scanning velocity (or shortest integration time per pixel) still compatible with a detection limit of $\sim 1$ Jy has been chosen. This made it possible to detect Sgr A* in every single image at the 3 different wavelengths where observations were made and thus allowed us to check the pointing in the individual maps before they were averaged to produce the final images. The largest shifts (up to 8″) were observed after sunrise (see Appendix B). The averaged – i.e., shifted and stacked – images are shown in Figs. 1a, b, c.

The planet Uranus served as the primary flux density calibrator (Orton et al. 1986; Griffin & Orton 1993). This planet could be observed at the same elevations as Sgr A* $\sim 1.5$ hours later on each night. Maps were made of the planet Uranus, both to calibrate the data, and to measure the telescope 'error-beam' (see Appendix A). Since the weather conditions were very stable, we estimate the accuracy of the flux densities derived for Sgr A* to be of order $\pm 15\%$, 30% and 30% at the wavelengths 800, 600 and 450 $\mu$m, respectively. The corrections for atmospheric extinction were applied by directly determining the 1.3 mm extinction and adopting opacity ratios of 7, 15 and 20 at the above wavelengths. These ratios were measured from photometric observations of Uranus, together with the simultaneous 230 GHz opacity measurements. High-airmass corrections for the altitude of Mauna Kea were applied. Because of image rotation the individual maps at each particular frequency overlap only in the central $\sim 70'' \times 50''$. The uncertainties given in Table 2 refer to this region. Towards the map edges the noise increases by factors of $2 - 3$.

The morphology of the dust emission in the central $\sim 90''$ is best recognized in the $\lambda 450\,\mu$m image Fig. 1c. A more extended $\lambda 450\,\mu$m map also obtained with the JCMT has been published by Dent et al. (1993). We refer to this latter paper for a more detailed discussion of the morphology of the dust emission and its correlation with molecular line emission. The emission from the inner part of the CND forms an elliptical ring of size $\sim 70'' \times 40''$ with its major axis oriented parallel to the Galactic plane (which extends in this image approximately from N–E to S–W). It encloses a 'flat valley' of low surface brightness – in the following referred to as Central Cavity – at the centre of which Sgr A* is located. The ringlike structure of the CND dust emission attains a minimum in the S–E direction, which is also visible in maps of molecular line emission such as HCN(1-0) (see, e.g., Güsten et al. 1987). The region of brightest [OI]$\lambda 63\,\mu$m emission observed by Jackson et al. (1993) which is located $\sim 25''$ N–E of Sgr A* is visible, e.g., in our $\lambda 450\,\mu$m image (Fig. 1c) as well as in the more extended map of Dent et al. as a weak dust emission feature. Most of the underlying extended emission from the East Core is not observed, due to the finite size of the maps and the dual-beam observing method used, so that the zero level of the dust emission shown in Fig. 1 has to be determined in a different way (see, e.g., Fig. 1 in M94). All submm dust emission maps confirm convincingly the suggestion by Davidson et al. (1992) of the existence of a Central Cavity of very low dust surface density of radius $R \sim 1$ pc (see also sect. 3.2 and 3.3).

As discussed in sect. 1 the surface brightness of dust emission increases $\propto \nu^4$ while that of Sgr A* is rather constant. This is demonstrated by the cuts shown in Figs. 1a–c which run diagonally (p.a. 45°, i.e., in direction SW–NE) through the images. In spite of the larger beam, Sgr A* is still the brightest source at $\lambda 800\,\mu$m whilst at $\lambda 450\,\mu$m the surface brightness of the lobe of the CND associated with the Western Arc (which does have a $\nu^4$ dust spectrum) has already increased to about twice the flux density of Sgr A* (see also M94). Flux densities of Sgr A* at the four wavelengths (including the IRAM-30 m telescope [MRT] observation described in the following section) have been obtained by fitting two-dimensional Gaussian functions to the central point source. The resulting flux densities are given in Table 2.

The revised spectrum is shown in Fig. 2a. Our observations are combined with flux densities obtained elsewhere (see Appendix D). At $\nu \lesssim 45$ GHz the estimated uncertainties are considerably smaller than the flux density variations observed as a function of time. Hence the symbol ○——○, which shows maximum and minimum observed flux densities, also represents a lower limit to the time variability.

The variability of the radio spectrum $S_\nu(A^*, 1)$ is well established. Zhao et al. (1991) suggest that for $\nu \geq 8.5\,\mathrm{GHz}$ intrinsic variations on time scales of a few months determine the variability. Between 1990.2 and 1991.2 three radio outbursts have been observed. Wright and Backer (1994) presented simultaneous observations of Sgr A* in the frequency range 1.4 to 89 GHz made in winter/spring 1990 in three time intervals of less than one month separation and during the decay of one of the outbursts. The observed variations are strongest (nearly a factor $\sim 10$) at $\lambda 3.4\,\mathrm{mm}$. However, the agreement between the correlated flux densities obtained as the result of two different VLBI experiments at different times (see Appendix D) are hard to reconcile with the lower of the two flux densities measured by Wright and Backer.

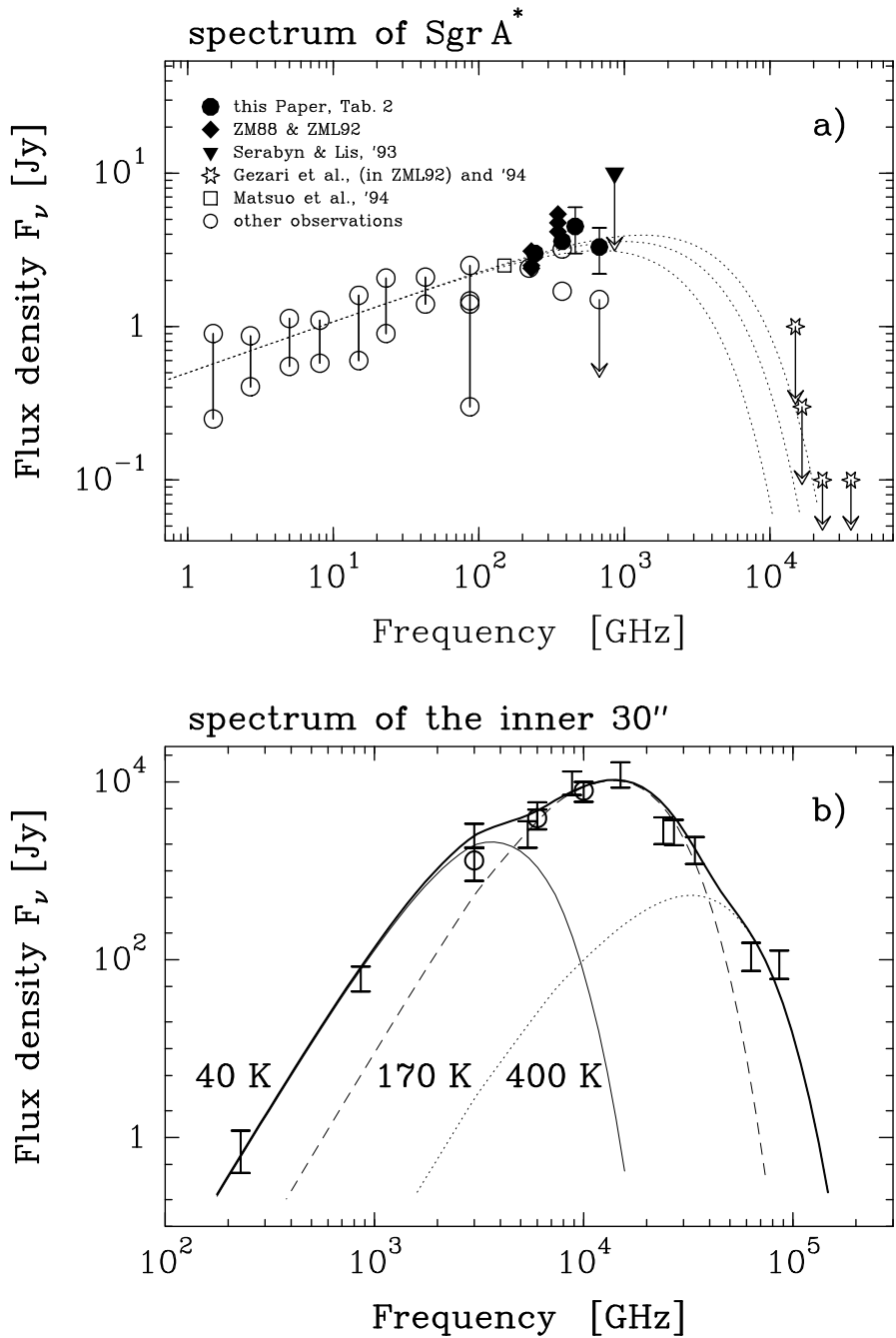

Figure 2: Spectral energy distribution a) The spectrum of Sgr A*. For $\nu \lesssim 100\,\mathrm{GHz}$ the highest and lowest observed flux densities are connected yielding the symbol ○——○ which indicates an upper limit to the amplitude of time variability. At frequencies $\nu \gtrsim 100\,\mathrm{GHz}$ the intrinsic uncertainties are considered to be greater than the observed variations. Therefore the individual observations are shown but not connected. Upper limits are indicated by downward directed arrows. The three dashed curves indicate spectra of optically thin synchrotron radiation from monoenergetic relativistic electrons. Model parameters have been selected for three cases consistent with observations. b) Dust emission spectrum of the central 30″, dereddened for a visual extinction of $A_v = 31\,\mathrm{mag}$. The spectrum has been decomposed into three components of different temperatures.

Table 2: Observed mm/submm flux densities of Sgr A*

| Date | $\lambda$ | $S_\nu$ | $\pm\Delta S_\nu$ |
|---|---|---|---|
| | $\mu$m | Jy | Jy |
| 08.-18.04.94 | 1300 | 3.3 | 0.3 |
| 27.02.94 | 800 | 3.5 | 0.5 |
| 01.03.94 | 600 | 4.0 | 1.2 |
| 27.-28.02.94 | 450 | 3.0 | 1.0 |

The upper limits in the MIR regime $18\,\mu\mathrm{m} \geq \lambda \geq 8\,\mu\mathrm{m}$ are due to the work by Gezari and coworkers (see ZML 92 for reference). A new upper limit of $\leq 1\,\mathrm{Jy}$ has recently been obtained at $\lambda 20\,\mu\mathrm{m}$ by Gezari et al. (1994a) which further constrains model fits to the spectrum $S_\nu(\mathrm{A}*,2)$. Dereddening could increase the upper limits by as much as a factor $\sim 2$ (see sect. 2.3).

## 2.2 Observations with the MRT: Variability of $S_\nu(\mathrm{A}*,2)$

The variability of $S_\nu(\mathrm{A}*,2)$ was investigated by ZML92 at $1300\,\mu\mathrm{m}$ and $870\,\mu\mathrm{m}$ on time scales of $\sim 1$ to 3 yr. No variations larger than $\sim 10-20\%$ at $1300\,\mu\mathrm{m}$ and $\sim 40\%$ at $870\,\mu\mathrm{m}$ were observed. Gwinn et al. (1991) investigated the short term variability of Sgr A* at the same wavelength with the CSO on time spans between 0.1 seconds and 24 hours, also with negative results.

The results reported here on time scales of $\sim 3$ days support the earlier findings. These observations were carried out at $\lambda 1300\,\mu\mathrm{m}$ with the IRAM 30-m telescope (MRT) during the bolometer campaign from 8 to 25 April 1994. The 7-channel MPIfR bolometer array (Kreysa et al. 1993) was used. The 7 channel passbands have each an equivalent bandwidth of $\sim 60\,\mathrm{GHz}$ and are centered at $\sim 240\,\mathrm{GHz}$. The effective beam shapes measured with Uranus, Mars and Neptune can be approximated by Gaussian functions of HPW $11''$ (see Appendix A).

Daily observations were planned during the first week and observations every third day during the two following weeks. Due to varying weather conditions measurements were limited to 5 days. Good weather conditions for reliable variability measurements ($\tau_{240\,\mathrm{GHz}} \leq 0.15$, no visible effects of anomalous refraction) prevailed only during 3 days where the following observing sequence has been repeatedly applied to Sgr A* and subsequently to the calibrator in order to minimize the observational overheads: skydip at azimuth of the source – pointing – focussing – pointing (to account for possible effects of anomalous refraction) – cross scans of $40'' - 60''$ length. All observations were performed at elevations $\geq 18°$. Since calibrators were positioned very close to Sgr A* the derived flux densities of Sgr A* are believed to have uncertainties of less than 10%.

Results of these observations, combined with earlier results, are shown in Figs. 3a and b where flux densities at $\lambda 1300\,\mu\mathrm{m}$, $870\,\mu\mathrm{m}$ and $800\,\mu\mathrm{m}$ are plotted as function of time. Within the observational uncertainties no variability at these wavelengths has been detected although the $\lambda 870\,\mu\mathrm{m}$ data may indicate a systematic variation on time scales of a few years. However, the observed flux density variations are still smaller than the uncertainties due to the observations made with different telescopes. This is demonstrated by a comparison of the error bars given in Fig. 2a for our $\lambda 600$ and $450\,\mu\mathrm{m}$ JCMT observations with the variation of the observed

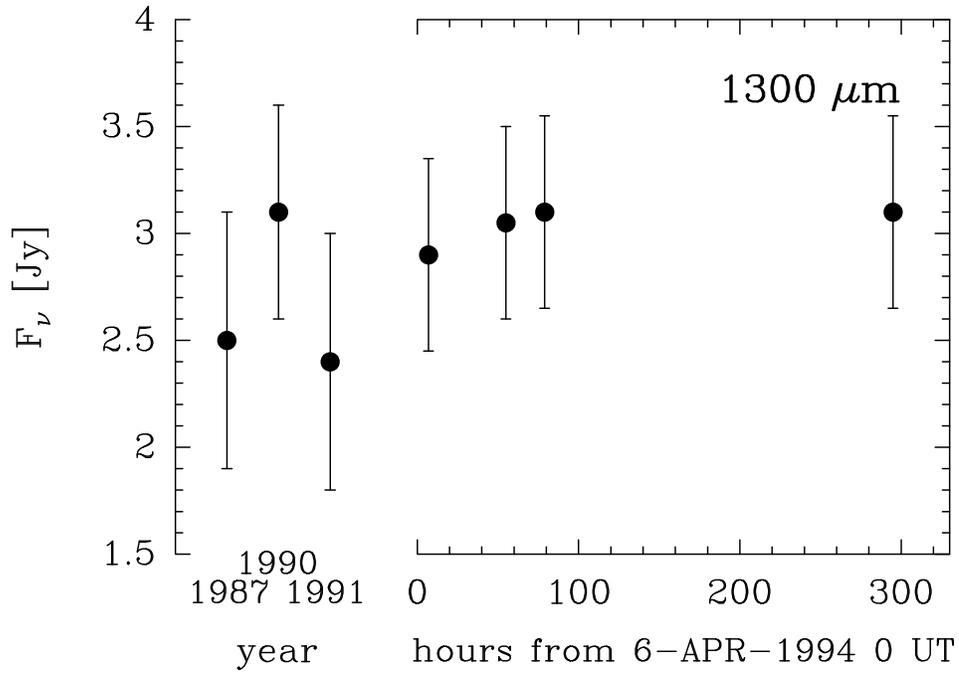

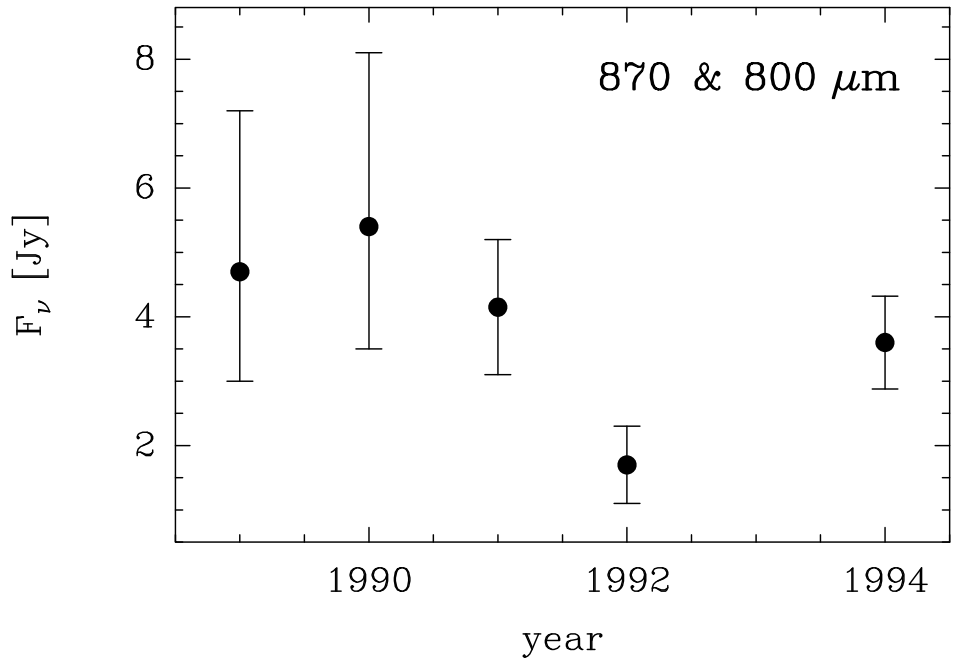

Figure 3: Flux densities of Sgr A* as function of time, measured a) at $\lambda 1.3$ mm with the MRT and b) at $\lambda 0.87$ mm and 0.8 mm, with MRT and JCMT respectively. Note the different time scales of the abscissa.

Table 3: Fit parameters for thermal dust emission from the central 30 arcsec

| $T_d$ | $\lambda_{\tau=1}$ | $N_H^{1)}$ | $\Theta_{eff}$ | $\Omega_{eff}$ | $M_H^{1,2)}$ | $L^{2)}$ |
|---|---|---|---|---|---|---|
| K | $\mu$m | cm$^{-2}$ | arcsec | arcsec$^2$ | M$_\odot$ | L$_\odot$ |
| 40 | 12 | 2.9E22 | 30 | 1020 | 393 | 2.1E5 |
| 170 | 0.8 | 4.3E21 | 15 | 255 | 3.9 | 4.3E6 |
| 400 | 0.02 | 4.8E19 | 3 | 10 | 6E-3 | 5.2E5 |

Footnotes:

[1] Computed for $Z/Z_\odot = 2, b = 1.9$

[2] Computed for $R_0 = 8.5$ kpc

MRT $\lambda 870\,\mu$m flux densities over several years. Further observations are planned to resolve this discrepancy.

## 2.3 The revised submm/IR spectrum of the inner 30 arcsec

In ZML 92 and M94 we have tried to estimate the luminosity of Sgr A* by determining and integrating the spectrum of the inner 30″. This procedure implies that there is enough dust in the inner pc (a visual extinction of $A_v \sim 2$ mag within 15″ of Sgr A* would be sufficient) to absorb all optical/UV emission from Sgr A*. Integration of the spectrum shown in ZML92 as Fig. 4b yielded $L_{IR}(30'') \sim 1.8\,10^6\,L_\odot$, of which a luminosity of only $< 10^6\,L_\odot$ could be attributed to Sgr A*.

More recently Gezari (1992) and Gezari et al. (1994b) obtained high-resolution MIR images at 8 different wavelengths ranging from $4.8\,\mu$m to $18.1\,\mu$m using a $58 \times 62$ pixel array camera system with the IRTF 3-m-telescope. Strong emission is confined to the central 30″ and coincides with the ionized gas of the Northern Arm and the East-West Bar as outlined by free-free radio emission. The transition from images dominated by reddened stars and circumstellar dust to images dominated by dust associated with interstellar neutral and ionized gas occurs between $4.8\,\mu$m and $7.8\,\mu$m. $\lambda 12.4\,\mu$m/$7.8\,\mu$m color temperatures range from 180 to 300 K for the interstellar dust while some of the IRS sources are considerably warmer. Gezari et al. derived model spectra, dust temperatures and opacities for every image pixel and corrected for absorption by interstellar dust using the strength of the $\lambda 9.8\,\mu$m silicate feature as an indicator. In this way a dereddened luminosity of $L_{IR} \sim 5\,10^6\,L_\odot$ was obtained which is nearly 3 times the luminosity estimated by ZML92 from observed flux densities which were not corrected for interstellar extinction.

To check if these two luminosities can be made consistent we corrected the spectrum Fig. 4b of ZML92 for extinction. Adopting $A_v \sim 31$ mag as the visual extinction between Galactic Center and Sun (Rieke et al. 1989), we computed the optical absorption depth $\tau_\lambda = (31/1.086)(A_\lambda/A_v)$ for each wavelength $\lambda$ where flux densities are given in Table 2b of ZML92. Relative extinction cross sections $(A_\lambda/A_v)$ were taken from Mathis et al. (1983) Appendix C, which in the range $\lambda 1.0 - 13\,\mu$m follow closely the extinction law determined for the Galactic Center by Rieke and Lebofsky (1985). The thus dereddened spectrum $S_\nu(30'')$ is shown in Fig. 2b. It has been decomposed into three components of dust temperatures 40, 170 and 400 K. Fit parameters are given in Table 3.

As to be expected dust temperatures are shifted to higher values and luminosities are increased relative to the values given in ZML92. The total dereddened luminosity is $L_{IR} \sim 5.0\,10^6\,L_\odot$, with $L_{IR}(\lambda < 30\,\mu$m$) \sim 4.2\,10^6\,L_\odot$, the latter value being in good agreement with the MIR luminosity derived by Gezari et al.. The fact that the dust temperatures of 170 and 400 K obtained from the model fit (Table 3) bracket the range of temperatures derived by Gezari (1992) confirms that emission from the Northern Arm and the East-West Bar dominate the IR luminosity in the central parsec.

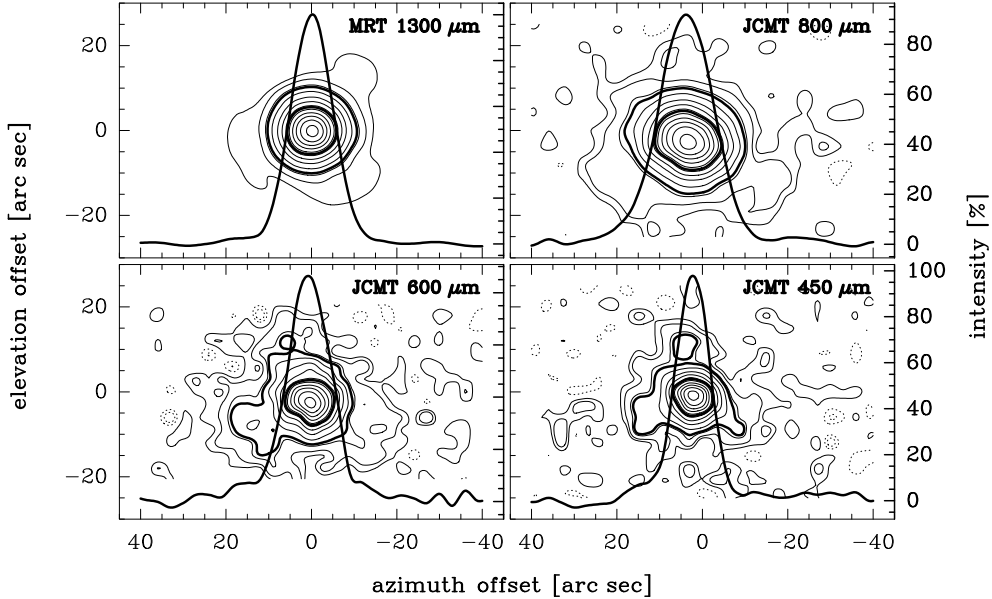

Figure 4: Uranus maps obtained with the MPIfR 7-channel bolometer array ($\lambda = 1300\,\mu$m) at the MRT and the UKT14 bolometer ($\lambda = 800$, 600 and $450\,\mu$m) at the JCMT. Contour levels increase from 2 to 4, 8 and 16% and from thereon in steps of 10% to 96%. Contours at 10% and 50% are indicated as heavy curves. Also shown are cuts in RA through the peak of the maps. The deconvolved HPBWs are $10\rlap{.}''8 \times 10\rlap{.}''5$ at $\lambda 1300\,\mu$m, $14\rlap{.}''4 \times 13\rlap{.}''4$ at $\lambda 800\,\mu$m, $10\rlap{.}''9 \times 9\rlap{.}''7$ at $\lambda 600\,\mu$m and $9\rlap{.}''3 \times 7\rlap{.}''9$ at $\lambda 450\,\mu$m.

The effective solid angles $\Omega_{\mathrm{eff}} = 1.133\,\Theta_{\mathrm{eff}}^2$ are chosen as follows: For the 40 K dust component $\Theta_{\mathrm{eff}} \sim 30''$ represents the telescope HPBW, for the 170 K dust component $\Theta_{\mathrm{eff}} \sim 15''$ represents roughly the area for which Gezari (1992) finds strong MIR emission (see specifically his Figs. 1 and 2). The $\lambda 12.4\,\mu$m opacity of this component which corresponds to $\lambda_{\tau=1} \sim 0.8\,\mu$m (Table 3) is $\tau_{12.4} \sim 0.04$, in reasonable agreement with the peak $12.4\,\mu$m emission opacity of $\sim 0.02$ derived by Gezari et al. (1994b). $\Theta_{\mathrm{eff}} \sim 3''$ for the 400 K dust component indicates that dust attains such high temperatures only in the immediate vicinity of the luminous dust embedded sources.

# 3   Discussion

## 3.1   Interpretation of the spectra (Sgr A*, 1+2)

Our new measurements at $\lambda 1300$, 800, 600 and $450\,\mu$m extend the observed spectrum to higher frequencies. The revised spectrum shown in Fig. 2a is qualitatively in agreement with the DL94 model described in sect. 1. The gap in the observed spectrum between $\lambda 450$ and $20\,\mu$m allows yet the presence of $\sim 3\,\mathrm{M}_\odot$ of 50 K dust and associated hydrogen gas within the telescope beam.

To the observed spectrum (Fig. 2a) we have fitted three spectra with the functional relationship $S_\nu \propto \nu^{1/3} \exp(-\nu/\nu_c)$ which are compatible with the time averaged observed flux densities and upper limits. Fit parameters are given in Table 4. DL94 interpret this spectrum as being optically thin synchrotron radiation from relativistic electrons with a monoenergetic electron distribution. How such an energy distribution can be produced has yet to be investigated in detail, but the general feasability has been demonstrated by Lesch and Pohl (1992). DL94 envisage a black hole of $\sim 10^6\,\mathrm{M}_\odot$ surrounded by an accretion disk with a radial mass inflow rate of $\sim 10^{-6}\,\mathrm{M}_\odot/\mathrm{yr}$ which produces a luminosity of $10^5\,\mathrm{L}_\odot$ and has a characteristic source size at $\lambda 3\,\mathrm{mm}$ of $\sim 3\,10^{13}\,\mathrm{cm}$ (Krichbaum et al. 1993, 1994). For a cut-off frequency $\nu_c \sim 3000\,\mathrm{GHz}$ a magnetic field strength of $\sim 6\,\mathrm{G}$ and an electron energy of $\sim 121\,\mathrm{MeV}$ are required. The physics of accretion disks yields an independent estimate of the magnetic field strength in the vicinity of the black hole. For equipartition between magnetic and thermal energy density the characteristic magnetic field of an accretion disk close to a black hole of $\sim 10^6\,\mathrm{M}_\odot$, i.e., within a few to a few tens of the Schwarzschild radius of $\sim 3\,10^{11}\,\mathrm{cm}$ is of order $\sim 10\,\mathrm{G}$, i.e., in reasonable agreement with the above value derived from the cut-off frequency. This suggests that the magnetosphere of the black hole is the place where electrons are accelerated and synchrotron emission originates. Although monoenergetic synchrotron emission fits the observed radio/IR spectrum well we want to reiterate that it is one – but not the only – successful model. Other models found in the literature are emission from a jet (Falcke et al. 1994) and spherical accretion (Melia 1994; Mastichiadis and Ozernoy 1994). A critical review of different models can be found in Duschl and Lesch (1994b).

Table 4: Fit Parameters for Synchrotron Emission

| $\nu_c$ | $\nu_{max}$ | $B$ | $E$ | $L$ |
|---------|-------------|-----|-----|-----|
| GHz | GHZ | G | MeV | $L_\odot$ |
| 2E3 | 6.7E2 | 5.1 | 111 | 2.5E2 |
| 3E3 | 1.0E3 | 5.6 | 121 | 4.3E2 |
| 4E3 | 1.3E3 | 6.0 | 127 | 6.3E2 |

## 3.2 The characteristics of the FIR emission

Cox and Laureijs (1989) determined from IRAS data the FIR luminosity integrated within the radius R in the plane of the Nuclear Bulge. The angular resolution of the IRAS maps was ~5-6arcmin.

$$L_{\mathrm{FIR}}/\mathrm{L}_\odot = \begin{cases} 1.7\,10^6\,(R/\mathrm{pc})^{1.2} & R \leq 115\,\mathrm{pc} \\ 7.0\,10^6\,(R/\mathrm{pc})^{0.9} & 115 \leq R/\mathrm{pc} \leq 300 \end{cases} \quad (1)$$

is an analytical aproximation of the functional relation shown as their Fig. 5, which yields for $R \sim 250\,\mathrm{pc}$ a total FIR luminosity of $10^9\,\mathrm{L}_\odot$ and thus accounts for $\sim 10\%$ of the total IR-luminosity of the Galaxy (Cox and Mezger 1989). The derived $\lambda 60/100\,\mu\mathrm{m}$ color temperature increases from $\sim 23\,\mathrm{K}$ at $R \sim 200\,\mathrm{pc}$ to $35\,\mathrm{K}$ at $R \sim 15\,\mathrm{pc}$. Cox and Mezger also estimate that MIR emission $\lambda < 30\,\mu\mathrm{m}$ adds $\sim 20\%$ to the total IR luminosity of the Galaxy.

Within $R \lesssim 250\,\mathrm{pc}$ Odenwald and Fazio (1984) detected altogether 47 FIR sources, 37 of which are considered to be located at the Galactic Center. Most of these sources are associated with Giant HII regions and hence are indicators of recent massive star formation. Cox and Mezger estimate that these sources contribute $\sim 10\%$ to the total IR-luminosity of $\sim 10^9\,\mathrm{L}_\odot$.

The most detailed recent discussion of the spatial distribution of FIR emission within the central 8 pc comes from Davidson et al. (1992). For an area of $\Delta\alpha \times \Delta\delta \sim 4 \times 8\,\mathrm{pc}$ they derive a luminosity $L_{\mathrm{IR}}(\lambda \geq 30\,\mu\mathrm{m}) \sim 5\,10^6\,\mathrm{L}_\odot$. Maps at $90\,\mu\mathrm{m}$ and $50\,\mu\mathrm{m}$ obtained with an angular resolution of $15'' \times 24''$ and $10'' \times 19''$, respectively, show two peaks separated by $\sim 2\,\mathrm{pc}$. However, while these two peaks are positioned symmetrically relative to Sgr A* at $90\,\mu\mathrm{m}$, the northern peak in the $50\,\mu\mathrm{m}$ map is shifted towards the position of Sgr A*. Davidson et al. modelled the observed images by a flared dust disk of inner radius $R \sim 0.9\,\mathrm{pc}$ and flare angles ranging from $25°$ to $45°$, which is inclined to the line-of-sight by an angle $i \sim 60°$. The brightest FIR emission comes from dust in the region $0.9 \leq R/\mathrm{pc} \leq 1.7$, i.e., the transition region between the neutral Circum-Nuclear Disk (CND) and the HII region Sgr A West, where ionized and atomic gas appear to be mixed. In particular Fig. 1c and the more extended $\lambda 450\,\mu\mathrm{m}$ dust emission map by Dent et al. (1993) shows that here the dust surface density attains a maximum while the region inside $R \lesssim 1\,pc$ forms the Central Cavity of very low dust surface density. Davidson et al. derive for this Cavity a FIR luminosity of $\sim 1.6\,10^6\,L_\odot$. To explain the asymmetry in the $\lambda 50\,\mu m$ map a 'point source' of relatively warm ($70 < T_d/K < 130$) dust had to be placed between the Northern and Eastern ionized Arms of Sgr A West. The model fit requires a dust mass of 1 to $6\,M_\odot$ depending on dust temperature.

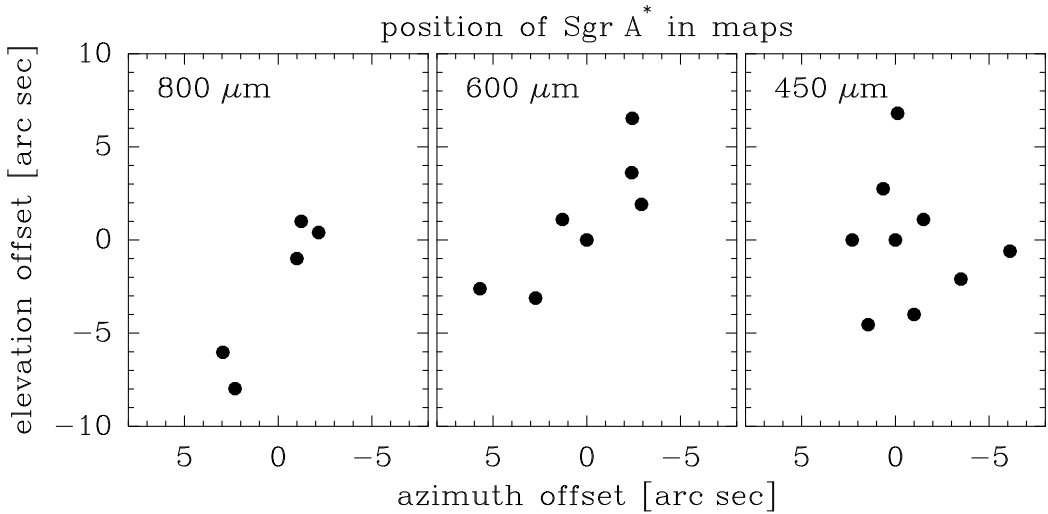

Figure 5: Observed positions of Sgr A* in the individual images relative to its nominal position. These shifts were corrected before the images (shown in Fig. 1) were averaged.

### 3.3 The characteristics of the MIR emission

In contrast to the FIR emission the MIR emission attains its highest surface brightness within $R \leq 0.6\,pc$. The high resolution maps and especially the extended mosaic $12.4\,\mu m$ map of Gezari (1992) show that the distribution of the MIR surface brightness is not symmetric relative to Sgr A* but is rather closely associated with the Northern Arm and the East-West Bar formed by the ionized gas of Sgr A West. The correlation between ionized gas and MIR emission has been discussed in detail by Gezari and Yusef-Zadeh (1990). A comparison with, e.g., the $\lambda 6\,cm$ VLA map by Lo and Clausen (1983) shows, however that the MIR surface brightness decreases more rapidly with distance R than that of the free-free emission.

Telesco et al. (in prep. and priv.comm.) using arrays with the IRTF produced MIR maps at wavelengths as long as $\lambda 30\,\mu m$. In their maps weak emission extends throughout most of the inner cavity. At shorter MIR wavelengths the distribution of surface brightnesses in the Telesco et al. maps exhibits the same asymmetry seen in the Gezari et al. maps. At $\lambda 30\,\mu m$, on the other hand, the surface brightness shows a much higher degree of symmetry relative to the center of the cavity indicating that a substantial fraction of the emission at this wavelength comes from dust which is not associated with the Northern Arm and the East-West Bar but rather with the Western Arc and quite generally with the inner edge of the CND.

Jackson et al. (1993) detected [O I] $\lambda 63\,\mu m$ emission from a 'Tongue' of hot atomic gas protruding between Northern and Eastern arm, whose center is located at a distance of $\sim 25''$ to the N–E of Sgr A*. Since there is a good correlation between the radial velocity profiles of the [O I] $\lambda 63\,\mu m$ and H76$\alpha$ emission, respectively, Jackson et al. suggest that the Northern Arm is just the ionized surface of this Tongue of neutral gas. For beams centered on maximum [O I] emission they derive masses $M_H$ and column densities $N_H$ of 200 and $70\,M_\odot$ and $5\,10^{21}$ and $10^{22}\,cm^{-2}$, respectively for a beam size of $55''$ and $22''$. The total mass of neutral gas within the central cavity is estimated to $\sim 300\,\mathrm{M}_\odot$. Actually, our spectral analysis of the dust emission from the central $30''$ yields a cold ($\sim 40\,\mathrm{K}$) dust component associated with $\sim 400\,\mathrm{M}_\odot$ of hydrogen (see Table 3) which one might be tempted to identify with the neutral gas detected by Jackson et al., which we, however, prefer to interpret as dust along the line-of-sight but outside the Central Cavity. The submm maps, especially the $\lambda 450\,\mu\mathrm{m}$ maps of Dent et al. (1993) and our Fig. 1c show a rather flat bottom throughout the Central Cavity. At the peak position of [O$\mathrm{I}$]$\lambda 63\,\mu\mathrm{m}$ emission and with an $8''$-beam the $\lambda 450\,\mu\mathrm{m}$ maps show at most an increase of $\sim 1 - 2\,\mathrm{Jy/beam}$, corresponding (for an assumed dust temperature of $\sim 70\,\mathrm{K}$) to $N_\mathrm{H} \sim 0.5 - 1\,10^{22}\,\mathrm{cm}^{-2}$, just consistent with the column densities derived by Jackson et al. for their much larger beams. This could mean that the distribution of gas in the 'Tongue' is rather smooth or – should the gas be clumped – that the clump sizes are $\ll 8''$. It does not appear, however, that the Tongue with this column density extends all the way to Sgr A*.

Both Jackson et al. (1993) and Davidson et al. (1992) hypothesize that the 'point source' required by the modelling of the FIR dust emission coincides with this Tongue. The new results presented here and by Gezari (1992) show that this identification cannot be correct for the following reasons: i) The velocity-integrated [O$\mathrm{I}$] $63\,\mu\mathrm{m}$ emission attains its maximum at a distance of $\sim 25''$ NE of Sgr A* where the MIR surface brightness has decreased to very low values. Hence, the warm H$\mathrm{I}$ gas is not associated with warm dust. ii) The Tongue cannot coincide with the point source, since it is extended even in the FIR beams. iii) The model fit by Davidson et al. requires for the 'point source' flux densities of $1100 \pm 450\,\mathrm{Jy}$ at $\lambda 90\,\mu\mathrm{m}$, $2800 \pm 700\,\mathrm{Jy}$ at $\lambda 50\,\mu\mathrm{m}$ and $4000\,\mathrm{Jy}$ at $\lambda 30\,\mu\mathrm{m}$. Note, however, that the latter value is not the result of a model fitting but rather this flux density had to be added to the model to yield the flux density observed at $\lambda 30\,\mu\mathrm{m}$. In Fig. 2b the above model fit flux densities corrected for an interstellar extinction of $A_v = 31\,\mathrm{mag}$ are shown as open dots, which – at least for the shorter wavelengths – fit the observed dereddened spectrum very well. Hence the 'point source' required for the model fit must be identical with the extended ($\sim 15'' \times 15''$) region of high MIR surface brightness which containes $\sim 4\,\mathrm{M}_\odot$ of gas and dust (see Table 3) rather than with dust associated with the mass $M_\mathrm{H} \sim 200\,\mathrm{M}_\odot$ of atomic gas inferred for the Tongue by Jackson et al.

## 3.4 FIR/MIR luminosities

Based on the discussion in the two previous sections we give in Table 5 integrated luminosities as a function of galactic radius $R$ separately for the FIR ($\lambda \gtrsim 30\,\mu\mathrm{m}$) and MIR ($\lambda \lesssim 30\,\mu\mathrm{m}$) regions. There is good agreement between IRAS FIR luminosities and KAO FIR luminosities for $R \leq 1\,\mathrm{pc}$ but at larger radii the IRAS luminosities become larger, probably because the chopping mode used for the KAO observations becomes insensitive to the extended emission detected by IRAS.

From $R \sim 250\,\mathrm{pc}$ to $0.6\,\mathrm{pc}$ the observed dust luminosity is due to cold ($\sim 20 - 50\,\mathrm{K}$) dust emitting in the FIR. The KAO maps by Davidson et al. (1992) like the IRAS maps by Cox and Laureijs (1989) trace primarily dust in this temperature range. We estimate that in this region IR emission shortward of $\lambda 30\,\mu\mathrm{m}$ contributes $\sim 20\%$ to the total IR flux density (Cox and Mezger 1989) and correct $L_\mathrm{IR,tot}$ correspondingly. Warm ($\sim 200 - 400\,\mathrm{K}$) dust contributes significantly only within $R \lesssim 0.6\,\mathrm{pc}$ where MIR emission associated with the Northern Arm and the East-West Bar accounts for more than 80% of the total luminosity. The MIR contribution to the total IR luminosity has been underestimated in most of the earlier papers since no extinction correction has been applied. Specifically Davidson et al. derived IR luminosities from $\lambda 60\,\mu\mathrm{m}/90\,\mu\mathrm{m}$ color temperatures which attain a peak value of $\sim 70\,\mathrm{K}$ and a temperature as low as $\sim 40\,\mathrm{K}$ averaged over the inner 8 pc. From the spectrum Fig. 2b it is clear that the flux density at $\lambda 90\,\mu\mathrm{m}$ is dominated by relatively cold ($\sim 40\,\mathrm{K}$) dust emission and at $\lambda 60\,\mu\mathrm{m}$ by much warmer ($\sim 170\,\mathrm{K}$) dust emission and that therefore the $\lambda 60\,\mu\mathrm{m}/90\,\mu\mathrm{m}$ colour temperatures as well as the corresponding luminosities will be underestimated. The IR-luminosity of the Nuclear Bulge ($R \lesssim 250\,\mathrm{pc}$) accounts for $\sim 10\%$ of the luminosity of the Galaxy ($R \lesssim 12\,\mathrm{kpc}$) to which compact sources contribute $\sim 10 - 20\%$.

Table 5: Integrated IR luminosities and Lyc-photon absorption and production rates

| $R$ | $L_{FIR}$ ($\lambda > 30\,\mu m$) | | | $L_{MIR}$ ($\lambda < 30\,\mu m$) | | $L_{IR,tot}$ | Sources | $N'_{Lyc}$ | $N_{Lyc}$ | IRE |
|---|---|---|---|---|---|---|---|---|---|---|
| | 1) | 2) | 3) | 3) | 4) | 5) | 6) | 7) | 7) | 8) |
| pc | Solar luminosities $L_\odot$ | | | | | | % | s$^{-1}$ | s$^{-1}$ | |
| 0.5 | 7.4E5 | 7.0E5 | 8.0E5 | 4.2E6 | 5.0E6 | 4.9E6 | | | | |
| 1 | 1.7E6 | 1.6E6 | | | | 5.8E6 | | 3E50 | 1.3E51 | 4.5 |
| 4 | 9.0E6 | 5.6E6 | | | | 1.3E7 | | | | |
| 2.5E2 | 1.0E9 | | | | | 1.3E9 | 10 | 6.4E51 | 2.0E52 | 47.5 |
| 1.2E4 | 9.7E9 | | | | | 1.2E10 | 21 | 1.3E53 | 2.2E53 | 21.6 |

Footnotes:

1)   Cox and Laureijs (1989); for $R = 12$ kpc Cox and Mezger (1989)

2)   Davidson et al. (1992), adjusted for $R_0 = 8.5$ kpc

3)   This paper

4)   Gezari et al. (1994b)

5)   $L_{FIR}(1) + 4.2\,10^6\,L_\odot$ for $R \lesssim 4$ pc, this paper;
     $1.25 \cdot L_{FIR}(1)$ for $R \lesssim 250$ pc, Cox and Mezger (1989)

6)   Cox and Mezger (1989)

7)   $R \lesssim 1$ pc: this paper; $R \lesssim 250$ pc: Mezger and Pauls (1978);
     $R \lesssim 12$ kpc: Güsten and Mezger (1983), both corrected for $R_0 = 8.5$ kpc

8)   This paper, eq.(2)

## 3.5   The Lyc-photon production rate and the IR-Excess

The number $N'_{Lyc}$ of Lyman continuum photons absorbed per second by interstellar gas can be estimated from the free-free emission at a frequency where it is optically thin. With a flux density of Sgr A West of 21 Jy at $\nu = 106$ GHz (Mezger et al. 1989) and for $T_e \sim 6000$ K the number of Lyc-photons absorbed by the gas within the Central Cavity is $N'_{Lyc} \sim 3\,10^{50}$ s$^{-1}$ and the mass of ionized hydrogen is $M_{HII} \sim \Phi^{0.5}\,197\,M_\odot$. Here $\Phi = \langle n_e^2 \rangle / n_e^2$ is the volume filling factor, $\langle n_e^2 \rangle^{0.5} \sim 2.5\,10^3$ cm$^{-3}$ the volume averaged and $n_e \sim 10^4$ cm$^{-3}$ the local electron density in the central arms (see, e.g., the discussion in Jackson et al. 1993). Hence the mass of ionized gas contained in the arms of the minispiral should be $M_{HII} \sim 50\,M_\odot$.

An important characteristic of an HII region is its IR-excess, i.e., the IR-luminosity of the dust associated with the ionized gas expressed in terms of energy contained in Ly-alpha photons, which is

$$
\begin{aligned}
\text{IRE} &= \frac{L_{IR}}{N'_{Lyc}\,h\nu_\alpha} = 2.34 \left[ \frac{L_{IR}}{10^6\,L_\odot} \right] \left[ \frac{N'_{Lyc}}{10^{50}\,s^{-1}} \right]^{-1} \\
&= 1 + \frac{1-f}{f}\,\frac{<E_{Lyc}>}{h\nu_\alpha} + \frac{L_{\lambda > 912}}{N_{Lyc}\,h\nu_\alpha}\,\frac{(1 - e^{-\tau_u})}{f}
\end{aligned}
\tag{2}
$$

Here $f = N'_{Lyc}/N_{Lyc}$, the fraction of all emitted Lyc photons which are directly absorbed by dust, is a function of the effective Lyc absorption depth $\tau_{Lyc} \cdot \langle E_{Lyc} \rangle = L_{\lambda < 912}/N_{Lyc}$ is the average energy per Lyc-photon, $L_{\lambda < 912}$ and $L_{\lambda > 912}$ are the luminosity of the ionizing star(s) shortward and longward of the Lyc limit at $\lambda = 912\,\mathring{A}$ and $\tau_u$ is the optical dust depth for UV/optical photons. These quantities are given in Mezger et al. (1974) as a function of the effective temperature $T_{eff}$ of the ionizing star(s).

Substitution of the above values $N'_{\rm Lyc}$ and $L_{\rm IR}$ for $R \sim 1$ pc yields IRE $\sim 4.5$, a value which is typical for giant galactic HII regions such as Sgr B2 (see below) which are ionized by O5-O9 stars. However, if we treat the region of high MIR emission (which accounts for $\sim 80\%$ of $L_{\rm IR}$ but for only $\sim 20\%$ of $N'_{\rm Lyc}$) and the remainder of the Central Cavity separately another picture emerges: IRE$_{\rm MIR} \sim 17$ and IRE$_{\rm cav} \sim 1$. The large value is typical for an HII region ionized by stars with $T_{\rm eff} \sim 3 - 3.5\,10^4$ K which is opaque for Lyc and UV photons, i.e., $e^{-\tau_{\rm Lyc}}$, $e^{-\tau_{\rm u}} \ll 1$. The small value can be explained by a virtually dust-free HII region, where all Lyc-photons are absorbed by the gas but with a negligible dust depth $\tau_{\rm u}$ so that $e^{-\tau_{\rm u}} \sim 1$.

For the central 500 pc free-free emission and IRE have been investigated by Mezger and Pauls (1978). As in the case of the FIR emission the free-free emission consists of a number of point sources superimposed on an extended background. Most of the point sources are Giant HII regions ionized by recently formed O-stars, the most luminous of which is Sgr B2 with $L_{\rm IR} \sim 3.6\,10^6\,{\rm L}_\odot$, $N'_{\rm Lyc} \sim 3.5\,10^{50}\,{\rm s}^{-1}$ and IRE $\sim 3.8$ (Gordon et al. 1993). Extended Low Density (ELD) HII regions must account for the background emission. Mezger and Pauls find that for $R > 10$ pc $N'_{\rm Lyc}(R)$ increases in a similar way as the FIR luminosity (see sect. 3.2) and attains at $R \sim 250$ pc the value $6.4\,10^{51}\,{\rm s}^{-1}$. After correction for Lyc-photons lost by dust absorption or escape the total Lyc-photon production rate is estimated to be $N_{\rm Lyc} \sim 2\,10^{52}\,{\rm s}^{-1}$ or $\sim 10\%$ of Lyc-photon production rate in the Galactic Disk.

## 3.6 Origin of the excitation in the Galactic Center

### 3.6.1 The inner 8 pc

Within $R \lesssim 0.6$ pc (a) source(s) with $T_{\rm eff} \sim 3 - 3.5\,10^4$ K must be responsible for both the heating of the warm dust and the ionization of the gas inside the Central Cavity. If the heating were provided by a central source and the Cavity were virtually dust free the observed IR-luminosity would have to be multiplied by a geometrical factor of $\sim 2.5 - 5$ to account for photons not intercepted by the CND (see, e.g., Davidson et al. 1992)

Different arguments, however, speak against a central source of heating i) Dust color temperatures $\lambda 50/90\,\mu$m (Davidson et al.; see sect. 3.6.2) and at shorter wavelengths (Telesco, priv.comm.) decrease much more slowly ($\propto R^{-0.15}$ to $R^{-0.4}$) than would be expected if a central source provides the heating (in which case $T_{\rm d} \propto R^{-0.65\ldots-0.80}$ due to opacity effects); ii) On the basis of an independent argument Rieke, Rieke and Paul (1989) arrived at the same conclusion: if the heating were provided by a central source it should be easily detected as a NIR source – contrary to observations. Recently, however, a weak NIR source has been detected at the position of Sgr A*. From its dereddened NIR spectrum Eckart et al. (1993) estimate $\sim 1 - 5\,10^5\,{\rm L}_\odot$ as an upper limit of the luminosity of Sgr A*. iii) As has been pointed out by several authors there are enough massive and luminous stars within the central $30''$ to account for the heating of the dust and the ionization of the gas within the Cavity (see, e.g., the reviews by Genzel et al. 1994a,b and references therein). In fact the detailed investigation of Gezari (1992; see, e.g., his Fig. 10) indicates a clear correlation of dust temperatures with the positions of the HeI/HI stars detected by Krabbe et al. (1991). If these and other massive stars in the central pc have typical luminosities of $\sim 10^6\,{\rm L}_\odot$ they can heat dust to temperatures of $\sim 170$ K at distances of $\sim 5\,10^{17}$ cm. This corresponds to an angular distance of $\sim 4''$ which is typical for the half distance between the dust embedded sources in the MIR images of Gezari.

If stars are the sources of dust heating and ionization within $R \gtrsim 0.6$ pc we can estimate their total luminosity without invoking rather uncertain assumptions related to the geometry of gas and dust. The rather abrupt decrease of the MIR surface brightness at distances $R \gtrsim 15''$ indicates that the volume density of these luminous and hot stars must decrease steeply beyond $R \sim 0.6$ pc. While there are still enough Lyc-photons to keep gas in the Minispiral and at the inner edge of the CND ionized, the stellar radiation density appears to be too low to heat dust to temperatures above 200 K. Within this radius the luminous stars appear to be distributed rather uniformly (see, e.g., Genzel et al. 1994a,b, and references therein). Hence, the asymmetry in the distribution of the MIR surface brightness at $\lambda \lesssim 18\,\mu$m, where the Northern Arm and the East-West Bar are the most conspicuous objects (see sect. 3.3) must be due to an asymmetrical distribution of the dust rather than of the exciting stars. From the $\lambda 12.4\,\mu m$ map of Gezari we estimate the area of high MIR surface brightness to $\sim (15'')^2$ and the volume occupied by hot dust to $\sim (15'')^3$. We assume that most of the radiation of the luminous stars in this volume is absorbed by dust and that the stars are uniformly distributed within the central sphere of radius $R \sim 15''$. Hence the total luminosity of all stars within this sphere will be $L_* \sim 4\pi/3 L_{MIR} \sim 1.8\,10^7\,L_\odot$ and $N_{Lyc} \sim 4\pi/3 N'_{Lyc} = 1.3\,10^{51}\,s^{-1}$, respectively, with $L_{MIR}$ and $N'_{Lyc}$ taken from Table 5. Note that these luminosities produced in a volume of size $\sim 1-2\,pc$ are extreme but not unique. Similar physical conditions are found in the massive star forming cores of W49A/51A (Sievers et al. 1991) and – to a lesser extent – in Sgr B2 (Gordon et al. 1993).

In the CND, at $R \gtrsim 1-1.5\,pc$ stars with considerably lower $T_{eff}$ must be the dominant heating sources. Dust temperatures are below $\sim 50\,K$ and hydrogen appears to be predominantly neutral.

### 3.6.2 The Nuclear Bulge

Although the IRAS $\lambda 25$ and $12.5\,\mu m$ maps still show extended emission within the inner $1°$ the luminosity appears to be dominated by FIR emission. The $\lambda 60/100\,\mu m$ color temperature decreases from $\sim 35\,K$ at $R \sim 15\,pc$ to $\sim 23\,K$ at $200\,pc$ according to $\left(\frac{T}{35\,K}\right) \sim \left(\frac{R}{15\,pc}\right)^{-0.15}$ from whereon it stays constant. (Note, however, that Cox and Laureijs quote a functional dependence of $T_d \propto R^{-0.3}$ in contradiction to their Fig. 4a.) The work by Davidson et al. indicates that the above relation also holds for $R < 15\,pc$ yielding $T \sim 40\,K$ for $R \sim 8\,pc$ and $T \sim 50\,K$ for $R \sim 1\,pc$ in accordance with their Fig. 8a.

The total FIR luminosity inside $R \sim 250\,pc$ is $1.3\,10^9\,L_\odot$. This is $\sim 1/3$ of the estimated total luminosity of the stars in the Nuclear Bulge. Since the functional relationship of dust luminosity with galactocentric radius $R$ is very similar to the luminosity (inferred from their NIR surface brightness) of an older population of medium-mass stars which is dominated by M- and K-giants, these stars are considered to be the principal heating source of the dust in the Nuclear Bulge (Cox and Laureijs 1989).

What is the source of ionization outside Sgr A West? In Table 5 the (observationally determined) Lyc-photon absorption ($N'_{Lyc}$) and (estimated) production rates ($N_{Lyc}$) are given for the Nuclear Bulge ($R \lesssim 250\,pc$) and the Galaxy ($R \lesssim 12\,kpc$). The Nuclear Bulge contributes $\sim 12\%$ to the total IR-luminosity and Lyc-photon production rate. Since molecular hydrogen in the Nuclear Bulge ($\sim 10^8\,M_\odot$) also accounts for $\sim 10\%$ of the total mass of molecular hydrogen ($\sim 10^9\,M_\odot$) in the Galaxy we conclude that the rate of massive star formation per unit $H_2$ mass is similar in Bulge and Galactic Disk, although the larger value of the IRE may indicate that fewer early-type O-stars contribute to the ionization in the Bulge. As in the Disk medium-mass stars must be the main contributors to the heating of the dust (Cox and Mezger 1989, and references therein).

## 4  Summary and Conclusions

### 4.1  New results

In this paper we present the following results:

i) Submm dust emission maps at $\lambda 800$, 600 and $450\,\mu m$ of Sgr A* and the Central Cavity have been obtained with the JCMT. The dust column density in the Cavity is low, $\sim 1/5$–th to $1/10$-th that of the inner edge of the CND. A slight increase in the dust emission $\sim 25''$ N-E of Sgr A* is consistent with a hydrogen column density of $0.5 - 1\,10^{22}\,cm^{-2}$, inferred by Jackson et al. (1993) from [OI]$\lambda 63\,\mu m$ observations for a "Tongue" of $\sim 200\,M_\odot$ of neutral atomic hydrogen protruding from the inner edge of the CND towards the center of the Cavity. We estimate a mass of $\sim 50\,M_\odot$ of ionized hydrogen to be contained in the minispiral.

ii) The dereddened FIR/MIR spectrum (Fig. 2b) of the central $30''$ ($R \sim 0.6\,pc$) indicates that MIR emission from warm ($\sim 200-400\,K$) dust accounts for $\sim 80\%$ of the luminosity. MIR maps by Gezari (1992) show that this warm dust is associated with the Northern Arm and the East-West-Bar of the ionized Minispiral in the Central Cavity, but that its surface brightness decreases rapidly beyond $R \lesssim 0.6\,\mathrm{pc}$.

iii) Heating of the dust and ionization of the gas requires a source of excitation with $< T_{\mathrm{eff}} > \sim 3 - 3.5\,10^4\,\mathrm{K}$, a luminosity of $L_* \sim 2\,10^7\,\mathrm{L_\odot}$ and a Lyc-Photon production rate of $N_{\mathrm{Lyc}} \sim 1.3\,10^{51}\,\mathrm{s^{-1}}$. Evidence is given that this excitation can not be provided by a central source but rather by a cluster of hot and luminous stars which are distributed over the central pc.

iv) The revised spectrum of Sgr A* is shown in Fig. 2a. While at $\nu \lesssim 100\,\mathrm{GHz}$ there is clear evidence for variation on time scales of a few months the evidence for similar time variations in the mm/submm regime is marginal. The time-averaged spectrum can be approximated by optically thin synchrotron emission from "monoenergetic" relativistic electrons (shown as dotted curves in Fig. 2a). The gap in the observed spectrum could still accomodate within the telescope beam emission from $\sim 50\,\mathrm{K}$ dust associated with $\sim 3\,\mathrm{M_\odot}$ of hydrogen.

v) The integrated Sgr A* spectrum $\lambda \gtrsim 10\,\mu\mathrm{m}$ has a luminosity of only $250 - 600\,\mathrm{L_\odot}$. The detection of a weak and possibly extended NIR source coincident with the radio position of Sgr A* has been confirmed (Herbst et al. 1993; Eckart et al. 1993; Close et al. 1994). The latter authors observed a source in K, H and J which coincides within $0''.2$ with the radio position of Sgr A* and which may be variable. However, the detection of a HeI line which coincides with this source (Eckart et al. as quoted by Close et al. 1994) sheds some doubt on the identification of this source with an accretion disk surrounding a $\sim 10^6\,\mathrm{M_\odot}$ black hole. The dereddened spectrum of this source allows an estimated maximum luminosity which ranges from $\sim 2\,10^2\,\mathrm{L_\odot}$ (Herbst et al. 1993) to $\lesssim 1 - 5\,10^5\,\mathrm{L_\odot}$ (Eckart et al. 1993) but at any rate can not account for the bulk of the at least hundred times higher luminosity and Lyc-photon production rate required for the excitation of the central pc.

vi) Recent $\lambda 7\,\mathrm{mm}$ and $3\,\mathrm{mm}$ VLBI observations (Krichbaum et al. 1993; M94; Rogers et al. 1994) yield source sizes $\sim 0.33 - 0.13\,\mathrm{mas}$ ($\hat{=} 4 - 1.6\,10^{13}\,\mathrm{cm}$) as compared to a size of $2R_{\mathrm{S}} \sim 1.2\,10^{12}\,\mathrm{cm}$ for a black hole of $\sim 2\,10^6\,\mathrm{M_\odot}$ ($R_{\mathrm{S}}$: Schwarzschild radius). The corresponding brightness temperatures are $T_{\mathrm{b}} \sim 2\,10^9 - 1.4\,10^{10}\,\mathrm{K}$. The elongated structure detected by Krichbaum et al. has not yet been independently confirmed.

vii) Outside the central 2 pc but within the Nuclear Bulge ($R \lesssim 250\,\mathrm{pc}$) the ten times larger value of the Infrared Excess (IRE $\sim 47.5$ as compared to $\sim 4.5$ for $R \lesssim 1\,\mathrm{pc}$) signals a different source of excitation. A comparison of the indicators of massive star formation ($N'_{\mathrm{Lyc}}/M_{\mathrm{H_2}}$) and dust heating ($L_{\mathrm{IR}}/M_{\mathrm{H_2}}$), the latter being dominated by medium-mass stars (Mathis et al. 1983), shows an obvious similarity with the Galactic Disk, although fewer early-type O-stars may prevail in the Bulge.

## 4.2  The Galactic Center - a Laboratory for AGNs?

The dynamics of gas and stars in the central pc suggest the presence of a compact object of $\sim 2\,10^6\,\mathrm{M_\odot}$ (for a recent review see Genzel et al. 1994b). If it were an accreting black hole its maximum luminosity would be $L_{\mathrm{Edd}} \sim 6\,10^{10}\,\mathrm{L_\odot}$, far beyond the $\sim 2\,10^7\,\mathrm{L_\odot}$ inferred from FIR/MIR observations for a central luminous object. Falcke et al. (1993) have shown that a luminosity of $7\,10^4 - 7\,10^5\,\mathrm{L_\odot}$ and an effective temperature of $T_{\mathrm{eff}} \sim 2 - 4\,10^4\,\mathrm{K}$, which would be compatible with NIR observations, can be ascribed to a rapidly rotating Kerr black hole of $2\,10^6\,\mathrm{M_\odot}$, accreting $10^{-8.5} < \dot{M}/\mathrm{M_\odot}\,\mathrm{yr^{-1}} < 10^{-7}$ which is seen nearly edge-on.

Is Sgr A* a starved black hole as are supposed to exist in the centers of other normal galaxies (see, e.g., Kormendy 1994)? The configuration of a CND with a nearly empty Central Cavity, in which a Tongue of neutral gas and dust as well as the ionized Northern Arm and East-West Bar with an aligned magnetic field of $\sim 10\,\mathrm{mG}$ (Aitken et al. 1991) extend towards the center (i.e., Sgr A*) is suggestive. Observations and model computations supporting this scenario will be summarized and discussed in detail in a forthcoming review (Mezger, Duschl and Zylka, 1994). Here we want to address two specific problems, viz. the formation of the Central Cavity and the cause for massive star formation in the central pc, both of which appear to be a consequence of the presence of a massive black hole.

### 4.2.1 The Central Cavity

For a recent review of the CND see, e.g., Genzel (1988). It contains $\sim 10^4\,M_\odot$ of highly clumped gas and dust and extends to $R \sim 6\,pc$. At its inner edge ($R \sim 1\,pc$) the column densities of dust and gas attain a maximum; farther inward at least the dust column density drops by an order of magnitude.

As Duschl (1989) has shown this density structure is a consequence of the structure of the gravitational potential in the vicinity of the Galactic Center: At radii $R \lesssim 1\,pc$ it is dominated by a compact object of $\sim 2\,10^6\,M_\odot$, at larger radii it is determined by the mass distribution of the stars of the Nuclear Bulge whose volume density decreases $\propto R^{-1.8}$. This leads to a kink in the rotation curve and as a consequence of this, to a drastic change of the accretion time within a narrow radial range around this kink.

For larger radii the radial velocity is considerably smaller than the azimuthal velocity and thus allows the formation of the disk-like structure of the CND. For smaller radii the radial velocity is no longer negligible. Therefore the CND terminates at the radius where the radial velocity begins to increase.

This process not only explains the comparatively sharp inner edge of the CND but also the small amount of gas mass within the cavity. Even for a stationary mass flow rate, due to conservation of mass, the average matter density becomes much smaller inside the Cavity than in the CND.

### 4.2.2 The Origin of the Central Star Cluster

Within $R \gtrsim 0.6\,pc$ exists a cluster of hot, luminous stars whose presence can be explained as gravitational instability of an accretion disk surrounding a $10^6\,M_\odot$ black hole. We have calculated models of stationary viscous disks with radial mass flow rates between $10^{-6}$ and $10^{-7}\,M_\odot/yr$ and viscosity parameters $\alpha$ in the range between 0.01 and 1 (Shakura and Sunyaev 1973). We find that the enclosed disk mass within $10^3\,(10^4)\,R_S$ lies between 0.04 and 1.3 (1.0 and 32) $M_\odot$ just consistent with our upper limits (see sect. 3.1). Self-gravity becomes important for radii $\geq 1.5\ldots 5\,10^4\,R_S$, depending on the viscosity.

At larger radii this gravitational instability and hence fragmentation of the disk – followed by star formation – will continue. The formation of the cluster of HeI/HI stars about $10^6$ yr ago would have required of course a considerably more massive accretion disk with accordingly higher accretion rate and energy release.

Unexpectedly the model computations also led to a possible explanation of the observed time variations of the radio spectrum. Our models indicate that for radii smaller than $\sim 250\,R_S$ both a high and a low temperature solution for the vertical disk structure coexist which is usually the signature of an instability. The small radii at which the disk seems to be unstable allows the corresponding time scales to be short.

## Acknowledgements

The authors wish to thank P. Biermann, P. Cox, A. Eckart, H. Falcke, R. Genzel, D. Gezari, F. Melia, L. Ozernoy and Ch. Telesco for interesting discussions and the staff of the Joint Astronomy Centre, Hawaii, for their assistance while JCMT observations were carried out. The JCMT is operated by The Royal Observatories, UK, on behalf of the UK Particle Physics and Astronomy Research Council, the Canadian National Research Council and the Netherlands Council for Pur Research. DWT wishes to thank the Royal Observatories for funding a Senior Research Fellowship. WJD acknowledges support from the Bundesministerium für Forschung und Technologie through grant 05 2HD134 (Verbundforschung).

# Appendices

## Appendix A: Telescope beam shapes

The surface accuracy of both the MRT and the JCMT were improved during the winter of 1993/94. The JCMT dish was readjusted to improve the surface accuracy immediately before our observing run. We observed the telescope beam-shape of each telescope by mapping the planet Uranus. These maps were also used for calibration of the Sgr A* maps.

Altogether fifteen maps at 1.3 mm, two maps each at 800 and 600 $\mu$m, and four maps at 450 $\mu$m were obtained. Fig. 4. shows representative maps obtained with the MRT at 1.3 mm and the JCMT at 0.8, 0.6 and 0.45 mm in Fig. 4a and 4b-d respectively. Contours are drawn at the same level in all maps. The MRT image was obtained with the MPIfR 7-channel bolometer array. The chopper throw used was: 32 arcsec at 1.3 mm, 40 arcsec at 0.8 mm and 30 arcsec at 0.6 and 0.45 mm. The percentage of the chopped out error beam is thus comparable (see Appendix C). Note that the elongation of the beams in the scanning direction (azimuth) could be caused by an imperfect chopper waveform in combination with scanning in azimuth (see ZML92 Appendix A).

We derived the following beam dimensions (FWHM) in azimuth and elevation: $10.8'' \times 10.5''$, $14.4'' \times 13.4''$, $10.9'' \times 9.7''$ and $9.3'' \times 7.9''$ at 1.3, 0.8, 0.6 and 0.45 mm respectively. These are approximately as predicted by the diffraction limited beams of the two telescopes. The 0.8 mm images are clearly affected by atmospheric effects as they were observed during worse weather conditions (see Table 1). The low-level 'lobes' in the beam-shape observed at 600 and 450 $\mu$m, and to a lesser extent at 800 $\mu$m, were caused by slight inaccuracies in the elevation-dependent dish corrections being applied. These were subsequently corrected after our observing run. However, it was the Uraunus maps which we took during our run which we used to calibrate our data.

## Appendix B: Atmospheric effects and effective pointing accuracy

ZML92 discussed the atmospheric and instrumental effects on measurements with the IRAM 30 m MRT (altitude $\sim$ 2900 m). Here we present similar effects observed with the JCMT (altitude $\sim$ 4200 m). Fig. 5 shows the position shifts of Sgr A* relative to its nominal position. The telescope pointing was checked on NRAO530, IRAS 1629-24 and Uranus. The position of Sgr A* was determined in the restored maps. To estimate the shifts the 450 $\mu$m images were first smoothed to $10''$ to increase the signal-to-noise ratio. The 800 and 600 $\mu$m images were used with the original resolution. Sgr A* was visible in every single image and at all frequencies. It should be noted that Sgr A* would not have been visible in the averaged 450 $\mu$m image without these positional corrections. Only due to the applied shift was the contrast in the whole map sufficient to determine the flux density of Sgr A* with the quoted error of $\sim$ 30%. The shape of the western part of the CND matches that observed by Dent et al. (1993) in their single 450 $\mu$m map, so we believe that our pointing corrections are valid.

## Appendix C: Calibration errors

In the calibration of images obtained with broadband bolometers three principal error sources can be identified.

*Corrections of atmospheric extinction:* At MRT and JCMT atmospheric extinction is determined in different ways. At the MRT the extinction can be determined directly with the bolometer in use measuring the emission of the atmosphere at different airmasses (SKYDIP procedure). Calibration tests performed by the authors and the IRAM staff have shown that an accuracy in the range of 10% for elevations $> 25°$ and $\sim$ 15% for lower elevations is obtained. At the JCMT the atmospheric opacity is currently continuously monitored using the NRAO 'tipper' radiometer located at the adjacent CalTech Submm Observatory. Opacity ratios for the observed and the recorded frequency have to be measured. These ratios depend on the frequency and the atmospheric opacity itself.

*Effects due to broad-band observations:* The effective frequency is determined by the spectrum of the source and the transmission curves of both the focal plane filters as well as the spectral shape of the atmospheric windows. If – as is usually the case – planets are used as calibrators, their spectrum increases $\propto \nu^2$. The two extreme cases we have to deal with here are the spectra of optically thin dust emission ($S_\nu \propto \nu^4$) and the spectrum of Sgr A* ($S_\nu \propto \nu^{1/3}$). We have evaluated the corresponding integrals and found no variations of $\nu_{\mathrm{eff}}$ of more than 5% for all four bands used in our observations.

*Effects of the double-beam observing mode:* The double-beam observing method (Emerson et al. 1979) has been developed to eliminate the atmospheric effects. It works because these effects occur mainly in the near field of the telescope, where the two beams are nearly identical. Thus atmospheric emission seen by the 'positive' and 'negative' beam nearly cancels out. If the chop throw (the angular distance between the two beams) is smaller than the error beam of the telescope, the same also holds true for the error beams. We adopt in our observations very small chop throws and assume that the error beams cancel exactly.

## Appendix D: Flux densities of Sgr A*

The spectrum of Sgr A* as shown in Fig. 2a is based on the following observations: For $\nu \lesssim$ 100 GHz we use the compilation of data given and referenced in Duschl and Lesch (1994a). For $100 < \nu/\mathrm{GHz} \lesssim 800$ references to most flux densities are given in Fig. 2a. Additional flux densities, referred to as 'other observations' are given in Table 6.

Table 6: Additional flux densities of Sgr A* for frequencies > 300 GHz

| $\nu$ GHz | $\lambda$ $\mu$m | $S_\nu$ Jy | Footnote |
|---|---|---|---|
| 86 | 3750 | 1.47 | 1) |
| 86 | 3750 | 1.4 | 2) |
| 222 | 1350 | 2.4 | 3) |
| 374 | 800 | 1.7 | 4) |
| 374 | 800 | 3.2 | 5) |
| 674 | 450 | <1.5 | 6) |

Footnotes:
1) Krichbaum et al. (1994)
2) Rogers et al. (1994)
3) Serabyn et al. (1992)
4) This flux density has been estimated from the $\lambda 800\,\mu$m image by Dent et al. (1993). Note however a difference in the calibration: Their surface brightness of the Western Arc is only $\sim 70\%$ of the corresponding surface brightness in our $\lambda 800\,\mu$m image shown in Fig. 1a.
5) R. Hills, pers.comm. as quoted in M94.
6) Dent et al. (1993).

# References

Aitken D.K., Gezari D., Smith C.H., McCoughrean M., Roche P., 1991, ApJ 380, 419

Allen D.A., 1994, in: GH94

Altenhoff W.J., Baars J.W.M., Downes D., Wink J.E., 1987, A&A 184, 381

Close L.M., McCarthy D.W., Melia F., 1994, ApJ (in press)

Coulman C.E., 1991, A&A 251, 743

Cox P., Mezger P.G., 1989 A&AR 1, 49

Cox P., Laureijs R., 1989 Proc. IAU Symp.No.136 'The Center of the Galaxy' (M.Morris, ed.) p.121

Davidson J.A., Werner M.W., Wu X., Lester D.F., Harvey P.M., Joy M., Morris M., 1992, ApJ 387, 189

Dent W.R.F., Matthews H.E., Wade R., Duncan W.D., 1993, ApJ 410, 650

Duncan W.D., Robson E.I., Ade P.A.R., Griffin M.J., Sandell G., 1990, MNRAS 243, 126

Duschl W.J., 1989, MNRAS 240, 219

Duschl W., Lesch H., 1994a A&A 286, 431 (DL94)

Duschl W., Lesch H., 1994b, Proc. IAU Symp.169 (in press)

Eckart A., Genzel R., Hofmann R., Sams B.J., Tacconi-Garman L.E., 1993, ApJ 407, L77

Emerson D.T., Klein U., Haslam C.G.T., 1979, A&A 76, 92

Falcke H., Biermann P.L., Duschl W.J., Mezger P.G., 1993 A&A 270, 102

Falcke H., Mannheim K., Biermann P.L., 1994, A&A 278, L1

Genzel R.; 1988, Proc. IAU Symp.136 (M. Morris, ed.) p.393

Genzel R., Harris A.I., 1994, (eds.) The Nuclei of Normal Galaxies: Lessons from the Galactic Center, Kluwer Academic Publishers, Dordrecht, The Netherlands (in press) (GH94)

Genzel R., Hollenbach D., Townes C.H., 1994b, 'The Nucleus of our Galaxy' Reports on progress in Physics 57, 417

Genzel R., Hollenbach D.J., Townes C.H., Eckart A., Krabbe A., Lutz D., Najarro F.N., 1994c, in: GH94

Gezari D.Y., Yusef-Zadeh F., 1990, in 'Astrophysics with Infrared Arrays', AIP conference series 13, 214

Gezari D.Y., 1992, in 'The Center, Bulge and Disk of the Milky Way' (L.Blitz, ed.) Kluwer Academic Publishers, Dordrecht, The Netherlands, p.23

Gezari D.Y., Ozernoy L., Varosi F., McCreight C., Joyce R., 1994a, in: GH94

Gezari D.Y., Dwek E., Varosi F., 1994b, in: GH94

Gordon M.A., Berkermann U., Mezger P.G., Zylka R., Haslam G., Kreysa E., Sievers A., 1993, A&A 280, 208

Griffin M.J., Orton G.S., 1993, Icarus 105, 537

Güsten R., Mezger P.G., 1983, Vistas in Astronomy 26, 159

Güsten R., Genzel R., Wright M.C.H., Jaffe D.T., Stutzki J., Harris A.I., 1987, ApJ 318, 124

Gwinn C.R., Danen R.M., Middleditch J., Ozernoy L.M., Tran T.Kh, 1991 ApJ 381, L43

Herbst T.M., Beckwith S.V.W., Shure M., 1993 ApJ 411, L21

Jackson J.M., Geis N., Genzel R., Harris A.I., Madden S., Poglitsch A., Stacey G.J., Townes C.H., 1993, ApJ 402, 173

Kormendy J., 1994, in: GH94

Krabbe A., Genzel R., Drapatz S., Rotaciuc V., 1991, ApJ 382, L19

Kreysa E., Haller E.E., Gemünd H.-P., Haslam C.G.T., Lemke R., Sievers A.W., 1993, Conf.proc. of Fourth Intl. Symp. on Space Terahertz Technology, UCLA, Los Angeles, March 30-April 1, 1993

Krichbaum T.P., Zensus J.A., Witzel A., Mezger P.G., Standke K., Schalinski C.J., Alberdi A., Marcaide J.M., Zylka R., Rogers A.E.E., Booth R.S., Rönnang B.O., Colomer F., Bartel N., Shapiro I.I., 1993 A&A 274, L37

Krichbaum T.P., Schalinski C.J., Witzel A., Standke K., Graham D.A., Zensus J.A., 1994, in: GH94

Lesch H., Pohl M., 1992, A&A 264, 493

Lo K.Y., Claussen M.J., 1983, Nature 306, 647

Mastichiadis A., Ozernoy L.M., 1994, ApJ 426, 599

Mathis J.S., Mezger P.G., Panagia N., 1983, A&A 128, 212

Matsuo H., et al., 1994 (in prep.)

Melia F., 1994, ApJ 426, 577

Mezger P.G., Smith L.F., Churchwell E., 1974, A&A 268, 283

Mezger P.G., Pauls Th., 1978, Proc. IAU Symp. No.84 (B.Burton, ed.) p. 357

Mezger P.G., Zylka R., Salter C.J., Wink J.E., Chini R., Kreysa E., Tuffs R., 1989, A&A 209, 337

Mezger P.G., 1994, in: GH94 (M94)

Mezger P.G., Duschl W., Zylka R., 1994 A&AR (in prep.)

Odenwald S.F., Fazio G.G., 1984, ApJ 283, 601

Orton G.S., Griffin M.J., Ade P.A.R., Nolt I.G., Radostitz J.V., Robson E.I., Gear W.K., 1986, Icarus 67, 289

Rieke G.H., Lebofsky M.J., 1985, ApJ 288, 618

Rieke G.H., Rieke M.J., Paul A.E., 1989, ApJ 336, 752

Rieke G.H., Rieke M.J., 1994, in: GH94

Rogers A.E.E., Wright M.C.H., Bower G.C et al, 1994, ApJL (in press)

Serabyn E., Carlstrom J.E., Scoville N.Z., 1992, ApJ 401, L87

Serabyn E., Lis D., 1994, in: GH94

Shakura N.I., Sunyaev R.A., 1973, A&A 24, 337

Sievers A.W., Mezger P.G., Gordon M.A., Kreysa E., Haslam C.G.T., Lemke R., 1991, A&A 251, 231

Wright M.C.H., Backer D., 1994, ApJ 417, 560

Zhao J.-H., Goss W.M., Lo K.Y., Ekers R.D., 1991, in 'Relationships between Active Galactic Nuclei and Starburst Galaxies' (A.V. Filippenko, ed.) ASP Conference Series, Vol.31, p.295

Zylka R., Mezger P.G., 1988, A&A 190, L25

Zylka R., Mezger P.G., Lesch H., 1992, A&A 261,119 (ZML92)